# Natural rubber-clay nanocomposites: mechanical and structural properties

Camila A. Rezende[1,2], Fabio C. Bragança[2], Telma R. Doi[1,2], Lay-Theng Lee[1]*, Fernando Galembeck[2], François Boué[1]

[1] Laboratoire Léon Brillouin, UMR-12, CEA-Saclay, 91191 Gif-sur-Yvette Cedex, France

[2] Institute of Chemistry, University of Campinas, P.O. Box 6154, CEP 13083-970, Campinas-SP, Brazil

*Corresponding Author E-mail: Lay-Theng.Lee@cea.fr

## ABSTRACT

The mechanical properties of non-vulcanized natural rubber and dialyzed natural rubber-clay nanocomposites have been studied by uniaxial deformations to evaluate the reinforcement efficiency of the clay. We show that while non-rubber molecules contribute to auto-reinforcement, removal of these molecules improves significantly the performance of clay as reinforcement agent. These mechanical properties are discussed in relation to morphological aspects of the clay characterized by TEM and SANS. The nanocomposites prepared by "latex-mixing" with aqueous dispersions of clay are found to contain completely exfoliated clay lamellae in coexistence with tactoids. Improved mechanical properties of the nanocomposites can be modeled by the high aspect ratio of exfoliated clay platelets coupled with immobilized rubber matrix. Interestingly, presence of tactoids does not appear to compromise the excellent reinforcement properties of the exfoliated platelets. At high deformations, strain-induced alignment of the clay exhibits anisotropic scattering, with anisotropy increasing with clay concentration and stretching.

# INTRODUCTION

Recent research in composite materials containing polymeric matrix reinforced with inorganic fillers has concentrated on nanocomposites where the filler size is of nanometric scale and of the order of the characteristic length scale of the polymer. This reduced size, hence, increased specific surface area, distinguishes nanocomposites from conventional composites where filler particles are micron-size. As a consequence, lower charge-load is required to achieve required improved properties. The most common fillers include carbon black and silica and calcium carbonate. Following the success of Nylon 6-layered silicate nanocomposite developed at Toyota [1, 2], there has been an increase in the number of papers and patents on polymer-clay nanocomposites that report outstanding mechanical and optical properties, thermal resistance and low gas permeability [3, 4, 5]. In these applications, exfoliated clays are exploited for their combined properties of very high surface area (> 750 $m^2/g$) and high aspect ratio (> 100).

Improved mechanical properties such as tensile strength and elastic modulus of reinforced polymers depend on several factors: filler particle size and concentration, aspect ratio, dispersion and morphology (single versus aggregated particles or in the case of smectic clays, intercalated versus exfoliated) [6, 7, 8, 9, 10, 11]. All these factors lead to the concept of filler-network formation and a percolation threshold. In this respect, since spontaneous ordering varies inversely with aspect ratio of the particle [12], clay platelets make interesting reinforcing agents due to their high aspect ratio which is usually two orders of magnitude greater than spherical particles. Theoretically therefore, with exfoliated clay platelets, several-fold increase in elastic modulus can be achieved at very low clay volume fraction. In practice however, complete exfoliation is not easily obtained.

To exfoliate clay, two considerations are important: polymer-clay compatibility to ensure polymer chain penetration into the clay galleries, and, steric repulsion of the chains to push the platelets apart. Compatibility alone does not guarantee exfoliation since strong polymer-clay interactions can promote intercalation without exfoliation. Common methods that are employed to prepare polymer-clay nanocomposites therefore include modifications of the clay by exchange of the interlayer cation with an organic cation or by chemical modifications to form an organophilic clay for compatibility, and, shear forces during milling of clay-polymer melt for mixing and exfoliation [13, 14].

In this paper, we present studies of mechanical and structural properties of natural rubber-clay nanocomposites prepared using the "latex method". This method consists of mixing natural rubber latex (an aqueous dispersion of negatively charged rubber particles) and an aqueous dispersion of clay that is pre-swollen [15]. By taking advantage of water as an excellent swelling and exfoliating agent for the clay [16], chemical modification or organocation exchange is avoided. Thus, the physical chemistry of the components is controlled by solution and colloidal properties. The latex method was previously used to prepare natural rubber nanocomposites with smectic clay, presenting tensile properties similar to those of vulcanized rubber. Clay platelets were well dispersed and preferentially oriented in the matrix, resulting in translucent nanocomposite films with increased solvent resistance and swelling anisotropy [17]. Bragança et al [18] also employed this procedure to make nanocomposites of styrene-acrylic latex (poly (styrene-*co*-butyl acrylate-*co*- acrylic acid) latex particles prepared by emulsion polymerization) with Na-montmorillonite and its ion-exchanged derivatives containing Li, K or Ca. These results showed that improved mechanical properties in relation to the pristine latex (up to ten-fold increase in elastic modulus and two-fold in tensile strength) depend strongly on the counterion type. Using a quantitative model, it was shown that strong association between clay lamellae and rubber matrix due to electrostatic adhesion at the organic/inorganic interfaces may be as strong as covalent binding [19].

Natural rubber latex, apart from rubber hydrocarbon (polyisoprene), is also known to contain non-rubber materials, namely phospholipids, proteins and a host of inorganic ions, the most notable among them being metallic cations [20]. Although most of the protein molecules are removed by centrifugation during the latex purification, others remain strongly associated to the latex particles. Our past studies have shown that these molecules and charges accumulate in nanodomains, and calcium ions in particular, play important roles in gel formation, inorganic crystallite formation and interfacial compatibility [21, 22]. In this work, we explore the influence of these non-rubber components on the mechanical properties of natural rubber-clay nanocomposites by comparing non-dialyzed and dialyzed natural rubber. Mechanical properties are characterized by uniaxial deformation tests and structural properties of the clay are investigated using transmission electron microscopy and small angle neutron scattering.

**EXPERIMENTAL PROCEDURE**

**Materials**

Centrifuged natural rubber latex (NR) with 58.3% dry weight rubber (stabilized with 0.69% ammonia) provided by Orbys (São Paulo) was used in the preparation of the nanocomposites. Dialyzed natural rubber latex (DNR) was obtained by dialysis against $NH_4OH$ solution (1% v/v) using regenerated cellulose dialysis bags (SpectraPor MWCO 3.5K dalton) to eliminate excess ionic species and other small non-rubber molecules. The solution was changed daily until the conductivity was lower than 20 µS (8 days). The latex was then homogenized and filtered through a 150 µm sieve.

Sodium montmorillonite (MT) was acquired from Southern Clay Products (cation exchange capacity = 102 mequiv/100 g of clay) and used as received.

**Preparation of nanocomposites**

Clay was pre-dispersed in distilled water using a magnetic stirrer for 30 minutes. The dispersion was added progressively to the latex while stirring and the final mixture was stirred for another hour. The latex-clay mixture was then concentrated in a spinning evaporator (Rotavapor type) to remove water at 60 °C under reduced pressure. This produced a thick slurry that was cast onto a flat polystyrene rectangular mold (20 x 10 x 2 cm) and dried in an oven under air at 60°C (at least 48 hours). The concentrations of clay, latex and water used to prepare the nanocomposites are specified in Table 1.

Table 1. Concentrations of clay, latex and water used in the preparation of natural rubber-clay nanocomposites.

| Sample | Clay content (w/w) | Clay volume fraction ($\phi$)* | Clay (g) | Latex (g) | Water (mL) |
|---|---|---|---|---|---|
| NR0 | 0% | 0 | 0 | 20.0 | 20 |
| NR1 | 1% | 0.0032 | 0.2 | 19.8 | 20 |
| NR2 | 2% | 0.0065 | 0.4 | 19.6 | 40 |
| NR4 | 4% | 0.0131 | 0.8 | 19.2 | 100 |
| NR6 | 6% | 0.0199 | 1.2 | 18.8 | 120 |
| NR10 | 10% | 0.0342 | 2.0 | 18.0 | 160 |
| NR20 | 20% | 0.0737 | 4.0 | 16.0 | 220 |
| NR30 | 30% | 0.1200 | 6.0 | 14.0 | 250 |

* calculated using density of clay = 2.86 g/cm$^3$ as indicated by the manufacturer
* density of NR latex = 0.91 g/cm$^3$

**Characterization of nanocomposites**

<u>Mechanical properties</u>

Mechanical properties of the nanocomposites were evaluated by uniaxial stretching. Samples for the tensile tests were cut from cast films (thickness ~ 1 mm) in two different shapes: (1) dumbbell shape with dimensions recommended by the DIN 52504 standard (L = 40 mm; W = 4.5 mm) and (2) rectangular shape (L = 20mm; W = 10mm). Dumbbell-shaped samples are more appropriate for determining Young's modulus of the nanocomposite. However, due to limited distance covered by the stretching machine, dumbbell-shaped samples at low clay concentrations could not be stretched until rupture. The shorter rectangular-shaped samples were therefore used to obtain complete stress-strain curves to characterize maximum elongation and tensile strength. The rectangular-shaped samples also maintain sufficient width after stretching to be analyzed by small angle neutron scattering.

Tensile tests were carried out at room temperature using an MGW-Lauda stretching machine at constant strain rate (10 mm/min). The stretching force (F) was measured with a HBM Q11 force transducer and the force value was converted to stress ($\sigma$) unit using the relation: $\sigma = F(\lambda)\lambda/A_0$. $A_0$ is the cross section of the non-deformed film, $\lambda$ defines strain by $\lambda = L/L_0$ where $L_0$ and $L$ are the sample lengths before and after stretching, respectively. Multiplication of the numerator by $\lambda$ accounts for the decrease in cross section with increasing elongation. The stretching test was therefore independent of the exact value of the initial length. At least five tests were carried out for each sample and the results were averaged. Young's modulus was evaluated from the initial linear part of the stress-strain curves (Hooke's region).

<u>Transmission Electron Microscopy</u> (TEM)

For TEM analysis, ultra thin (ca. 50 nm) sections were cut normal to the plane of the nanocomposite film using a diamond knife (Drukker) in a Leica EM FC6 cryo-ultramicrotome cooled with liquid $N_2$ at -140 $^o$C. A Carl Zeiss CEM 902 (80 keV) transmission electron microscope was used to image the film. Images were acquired using a Slow Scan CCD camera (Proscan) and processed in the iTEM Universal Imaging Platform.

<u>Small Angle Neutron Scattering</u> (SANS)

SANS experiments were carried out on the spectrometer PAXE at the ORPHEE reactor (CEA-Saclay). Two different configurations were employed: neutron wavelength $\lambda = 6$ Å and sample-detector distance of 2 m, and $\lambda = 15$ Å and sample-detector distance of 5m. The resulting range of scattering wave vector, $q = 4\pi\sin\theta/\lambda$ covered by these configurations is 0.003 to 0.18 Å$^{-1}$. The first configuration is referred in this work as "high-q" and the second as "low-q".

Both unstretched and stretched samples were studied. Stretched samples were held in place by clamps attached to a vertical metal rod and taped to the sample holder. Scattering signals were collected by 2-D multidetectors. For isotropic scattering, the 2-D data were averaged radially to obtain iso-intensity scattering spectra, I(q) as a function of q. For anisotropic scattering, iso-intensity averaging of horizontal and vertical scattering was performed by considering sector areas ± 10° about the x-axis (horizontal) and ± 20° about the y-axis (vertical). Data treatment followed standard correction procedures using the expression [23, 24]:

$$I(q) = \frac{I_{S(f,T,t)} - I_{H(f,T;t)}}{I_{P(f,T,t)} - I_{h(f,T,t)}} - C \qquad \text{Equation 1}$$

where $I_S$, and $I_H$ are scattered intensities of the sample and empty sample holder, respectively; $I_P$ and $I_h$ are scattered intensities of PLEXI glass (1 mm) and empty PLEXI glass holder, respectively. These intensities are normalized by the incident neutron flux (f), transmission (T) and sample thickness (t). A constant electronic background (C = 1.2 10$^{-5}$) was subtracted from low-q curves of isotropic samples. This value was estimated using a cadmium block.

## RESULTS

**Mechanical Properties**

Elastic Modulus

Measurements of elastic modulus were carried out with standard dumbbell-shaped films. At least five tests were carried out for each sample; in all cases, good reproducibility was obtained indicating absence of large-scale inhomogeneity. Young's modulus values were extracted from the linear portion of the stress-strain curves at very small λ values (maximum deformation of 2%). For pure rubber matrix, the Young's modulus is slightly higher in NR compared to DNR sample but both values remain low, with $E_m$ < 1 MPa. For the nanocomposite films, Young's modulus increases with clay content, but the enhancement effect is significantly more pronounced for DNR where the elastic modulus increases by 2.5 times with the addition of only 1% clay and by two orders of magnitude for 10% clay.

Tensile Strength

Stress-strain curves for rubber-clay nanocomposite films stretched to rupture were obtained with rectangular-shaped samples that allow higher degree of stretching compared to standard dumbbell-shaped samples. Maximum elongation and strength at rupture can thus be evaluated from these films. In addition, these samples possess sufficient width after stretching to be studied by SANS. Figure 1 shows stress-strain curves for natural rubber (NR) and dialyzed natural rubber (DNR) films at different clay concentrations.

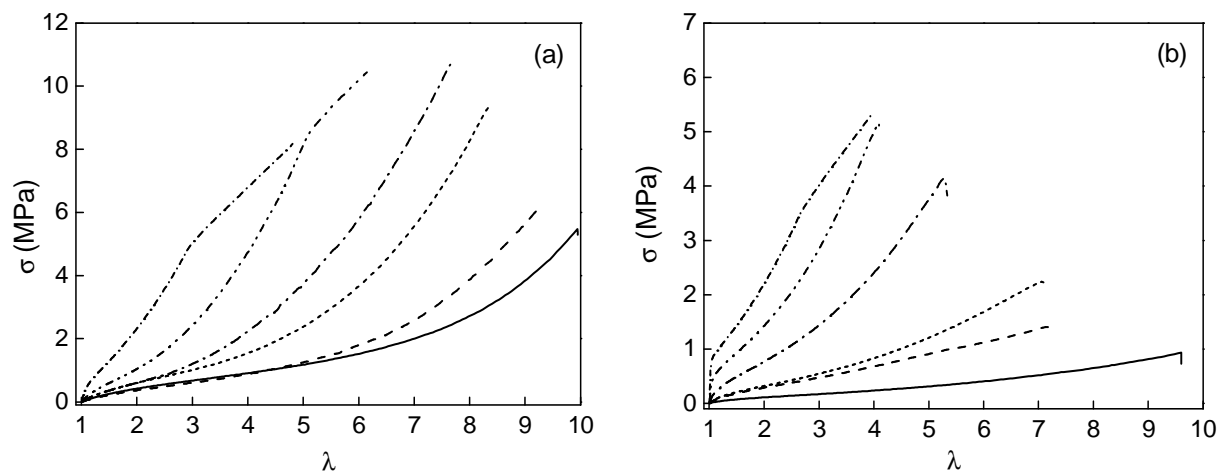

Figure1. Stress-strain curves of natural rubber-clay nanocomposites for (a) non-dialyzed rubber (NR) and (b) dialyzed rubber (DNR). The curves increase vertically with clay concentrations (by weight): 0% (solid line), 1% (long dash), 2% (dot), 4% (dash-dot), 6% (dash-dot-dot), 10 % (short dash).

The stress-strain profiles show low tensile stress at small deformations characteristic of elastomers with low reticulation degree, followed by an abrupt increase at high deformations. Note the surprisingly high tensile strength of pristine NR, where stress at rupture, $\sigma_{max}$ = 5-6 MPa. In the presence of clay, tensile strength increases with clay concentration. This increase in stiffness is accompanied by a corresponding decrease in maximum tensile strain, $\lambda_{max}$, at rupture (see Figure 2). Due to increasing stiffness with clay concentration and to the limitation of our stretching machine, only mechanical data of samples containing up to 10% clay (by weight) are considered reliable and presented here. The values for elastic modulus, maximum elongation ($\lambda_{max}$) and maximum tensile strength at rupture ($\sigma_{max}$) are given in Table 2.

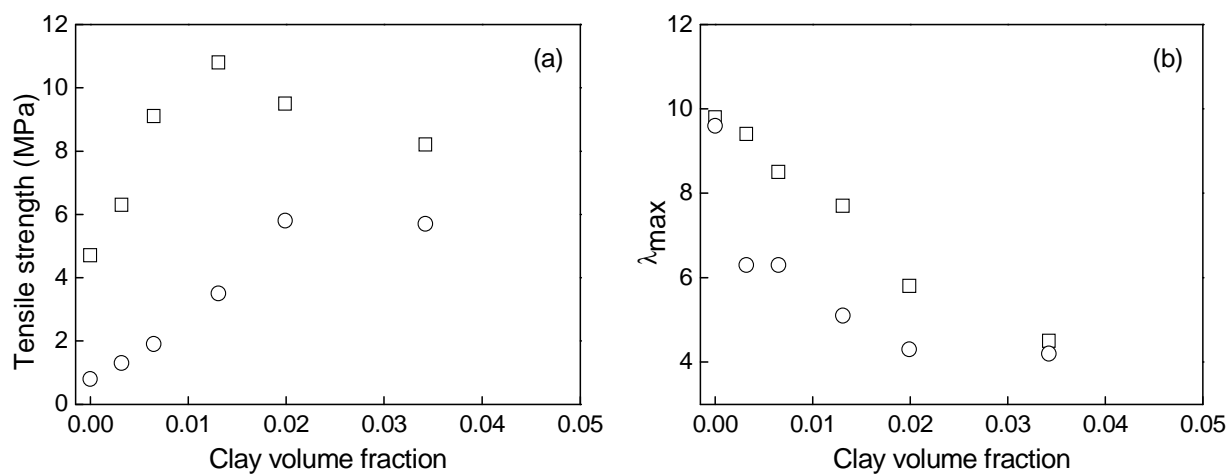

Figure 2. Effect of clay concentration on (a) tensile strength and (b) maximum elongation at rupture for NR (squares) and DNR (circles).

Table 2. Elastic modulus ($E_c$), maximum elongation ($\lambda_{max}$) and tensile strength ($\sigma_{max}$) of nanocomposites prepared from non-dialyzed (NR) and dialyzed natural rubber (DNR).

| Clay | Non-dialyzed rubber (NR) | | | Dialyzed rubber (DNR) | | |
| --- | --- | --- | --- | --- | --- | --- |
| Weight % | $E_c$ (MPa) | $\lambda_{max}$ | $\sigma_{max}$ (MPa) | $E_c$ (MPa) | $\lambda_{max}$ | $\sigma_{max}$ (MPa) |
| 0 | 0.9 ± 0.1 | 9.8 ± 1.0 | 4.7 ± 1.4 | 0.4 ± 0.1 | 9.6 ± 0.2 | 0.8 ± 0.2 |
| 1 | 0.9 ± 0.1 | 9.4 ± 1.0 | 6.3 ± 1.6 | 1.0 ± 0.1 | 6.3 ± 0.5 | 1.3 ± 0.1 |
| 2 | 1.4 ± 0.1 | 8.5 ± 0.1 | 9.1 ± 0.6 | 1.3 ± 0.1 | 6.3 ± 1.0 | 1.9 ± 0.6 |
| 4 | 1.7 ± 0.1 | 7.7 ± 0.5 | 10.8 ± 1.5 | 4.2 ± 0.6 | 5.1 ± 0.3 | 3.5 ± 0.6 |
| 6 | 3.5 ± 0.2 | 5.8 ± 0.7 | 9.5 ± 1.4 | 14.4 ± 2.2 | 4.3 ± 0.3 | 5.8 ± 0.8 |
| 10 | 4.9 ± 0.3 | 4.5 ± 0.4 | 8.2 ± 0.6 | 41.7 ± 3.0 | 4.2 ± 0.5 | 5.7 ± 0.8 |

<u>Reinforcement Factor</u>

Several theoretical models describing the mechanical behavior of filler-based nanocomposites have been presented in the literature [6, 7, 8, 9, 10, 11]. The earliest theory was based on Einstein's hydrodynamic theory for viscosity of colloidal suspensions. This theory was generalized by Guth and Gold [25], who computed pair-interaction coefficients to extend its application to higher concentrations. Later, it was applied to model elastic properties by replacing viscosity with elastic modulus, giving [26, 27]:

$$\frac{E_c}{E_0} = 1 + 2.5\phi + 14.1\phi^2 \qquad \text{Equation 2}$$

9$E_c$ and $E_0$ are Young's moduli of the nanocomposite and pure matrix, respectively, and $\phi$ is the volume fraction of filler particles dispersed in the rubber matrix. Equation 2 thus predicts reinforcement factor ($E_c/E_0$) to be dependent only on filler concentration and is valid only for spherical particles in the dilute regime. At higher concentration, it is found that the reinforcement factor increases more rapidly than is predicted by Equation 2. This is attributed to network formation or organization of filler particles into chain-like structures. To account for the "accelerated stiffening" caused by these chain-like or non-spherical fillers, Guth [27] introduced the shape factor $f$ in this equation, giving:

$$\frac{E_c}{E_0} = 1 + 0.67 f\phi + 1.62 f^2 \phi^2 \qquad \text{Equation 3}$$

Thus, for anisotropic particles with $f > 1$, Equation 3 (hereafter referred to as the Guth-Smallwood equation) assures a rapid increase in reinforcement factor with particle concentration.

For intrinsically anisotropic fillers such as rods and platelets, another widely used model is the Halpin-Tsai equation. Based on the self-consistent micromechanics methods developed by Hill [28] and Hermans [29], the Halpin-Tsai equation [30, 31] describes explicitly the significant effects of filler geometry alone on the composite mechanical properties. For oriented platelets, the reinforcement factor is expressed as:

$$\frac{E_c}{E_0} = \frac{(1 + 2f\eta\phi)}{(1 - \eta\phi)} \qquad \text{Equation 4}$$

where $f$ is the aspect ratio, $\eta = \dfrac{E_f/E_m - 1}{E_f/E_m + 2f}$ and $E_f$ the Young's modulus of the filler. For most inorganic fillers including clay, $E_f/E_0 \approx 10^5$ and $\eta \cong 1$.

The Halpin-Tsai model breaks down at high filler concentration, predicting lower values than obtained experimentally. Nielsen and Lewis [32, 33] modified this equation inserting a factor $\varphi$ to consider the maximum volumetric packing fraction of the filler:

$$\frac{E_c}{E_0} = \frac{(1 + 2f\eta\varphi)}{(1 - \varphi\eta\varphi)} \qquad \text{Equation 5}$$

where $\varphi = 1 + \phi[(1 - \phi_m)/\phi_m^2]$ and $\phi_m$ is the maximum volumetric packing fraction ($\phi_m$ = true volume of the filler/apparent volume occupied by the filler). The filler apparent volume is related to the matrix portion near the boundaries of the filler that are subjected to restricted mobility and thus contributes to reinforcement effect as if the filler had a greater volume. This is an important factor to be considered in clay nanocomposites due to the high specific surface area of the fillers and to polymer intercalated or trapped between the lamellae. Detailed calculations of the micromechanics of plate-shaped filler nanocomposites based on the above models are given in the paper by Bicerano et al [11].

In this study, we apply the models described by Equations 3 to 5 to fit the mechanical properties of the natural rubber-clay nanocomposites. The reinforcement factors are plotted as a function of clay volume fraction in Figure 3 for non-dialyzed and dialyzed rubber samples. In both cases, the reinforcement factor increases with clay concentration, with an initial slow-rise followed by a second rapid-rise regime. The best-fit curves using the Guth-Smallwood (Equation 3), the Halpin-Tsai (Equation 4) and modified Halpin-Tsai (Equation 5) models are also shown in the same Figure.

The Guth-Smallwood model uses only the aspect ratio as fitting parameter, and the fitted results obtained are: $f = 48$ and $f = 220$ for NR and DNR samples, respectively. The original Halpin-Tsai equation gives a reasonable fit at low filler concentration but as mentioned above, under-predicts the reinforcement factor at higher filler concentration. The modified form proposed by Nielsen and Lewis (Equation 5) on the other hand predicts the rapid-rise regime at increased filler concentration. Here, the two fitting parameters are: aspect ratio $f$ and maximum volumetric packing fraction, $\phi_m$ (true volume of filler/apparent volume occupied by filler). The best-fit curves yield $f = 65$ and $\phi_m = 0.048$ for NR, and $f = 300$ and $\phi_m = 0.038$ for DNR. Both models indicate a significantly higher degree of exfoliation of the clay in dialyzed rubber, and consequently higher reinforcement factor, compared to non-dialyzed samples.

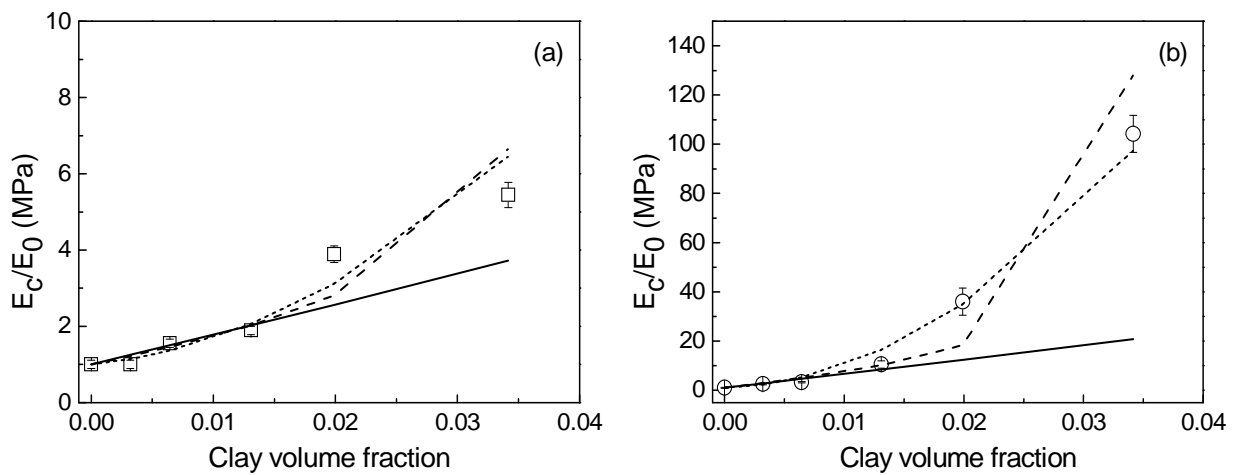

Figure 3. Reinforcement factor ($E_c/E_0$) versus clay volume fraction for (a) NR and (b) DNR nanocomposites. The lines are fits using Guth-Smallwood (short-dash), modified Tsai-Halpin (long-dash), and original Tsai-Halpin (solid) models.

**Clay Morphology Characterization**

The morphology of clay in the rubber matrix was characterized by electron microscopy and small angle neutron scattering. For non-dialyzed rubber films (in absence of clay), our previous studies show formation of calcium crystallites, ranging from several nanometers to several microns in size [22]. These large objects render image analysis of the clay platelets difficult. Therefore we restrict our studies to nanocomposites prepared from dialyzed samples.

<u>Transmission Electron Microscopy</u> (TEM)

Bright field TEM images of cross sections of DNR nanocomposite films containing 6% and 20% clay are shown in Figure 4. Clay lamellae (darker) appear distributed in the natural rubber matrix (clearer) in different degrees of exfoliation: individual clay platelets, partially exfoliated lamellae and tactoids constituted of packets of non-exfoliated clay platelets. Clay flexibility can also be seen in the amplified image in Figure 4c, as well as rubber-clay compatibility as evidenced by absence of holes and rupture at the polymer-clay interfaces.

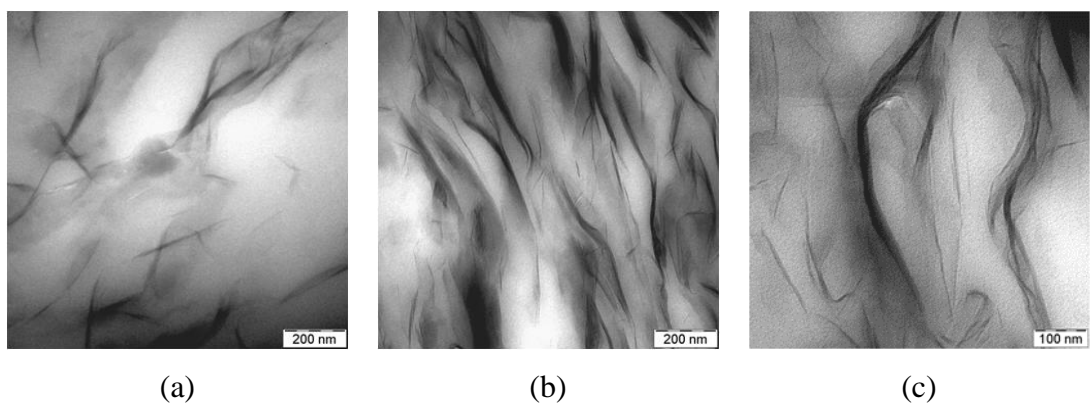

(a)          (b)          (c)

Figure 4. TEM of DNR nanocomposites containing (a) 6% and (b) 20% clay. The magnified image (c) shows co-existence of single clay platelets, partial exfoliates and tactoids.

Tactoid sizes were evaluated from more than 15 images for each nanocomposite totaling about 1200 domain counts for 6% (DNR6), and 1500 domain counts for 20% (DNR20). The size distributions are represented in histograms in Figure 5. For DNR6, the tactoid size ranges from about 2 to 20 nm (mean = 8.1 nm) and for DNR20, it ranges from about 2 to 35 nm (mean = 11.2 nm). From the tactoid size, the number of platelet/tactoid ($n$) can be evaluated from the relation:

$$t_{\text{tac}} = d(n-1) + e \qquad \qquad \text{Equation 6}$$

where $t_{tac}$ is the tactoid thickness, $d$ the interlamellar distance (without intercalation), and $e$ the lamella thickness. For the montmorillonite used in this study, $e = 0.92$ nm and $d = 0.96$ nm as determined by wide angle X-ray diffractometry (WAXD) [18]. The tactoid thicknesses thus give, $n \approx$ 2-21 platelets for DNR6, and $n \approx$ 2-36 platelets for DNR20; the corresponding average values are $n \approx 8$ and $n \approx 11$, respectively. Therefore, although the tactoid concentration increases with clay content, the average tactoid size does not vary significantly. It is to be pointed out that in this technique, imprecision in determination of tactoid thickness is introduced by the different orientations of the tactoid with respect to the electron beam.

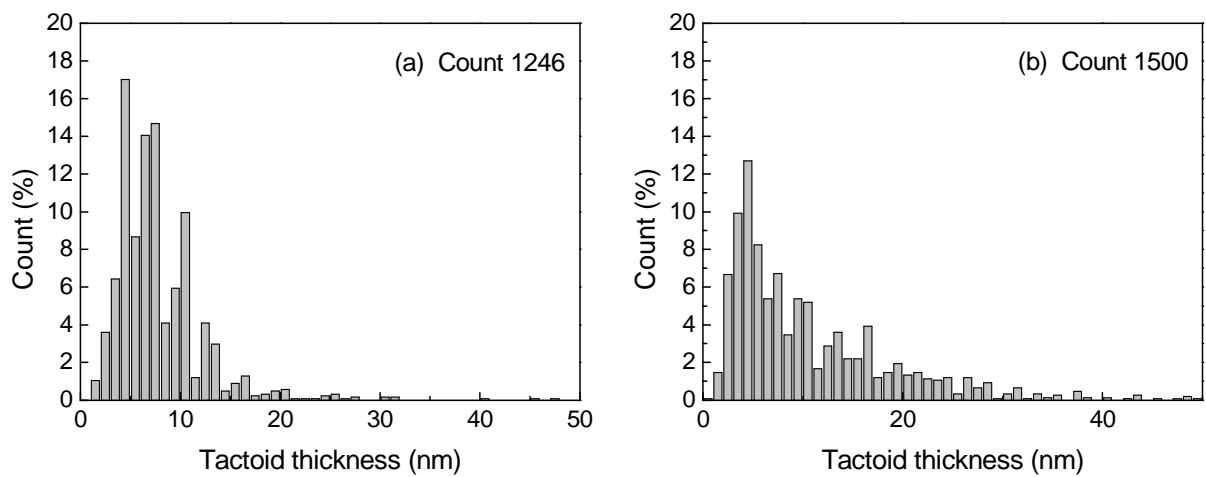

Figure 5. Histograms of tactoid size obtained by measuring transversal cross sections of clay domains in DNR nanocomposites containing (a) 6% and (b) 20% clay.

Small Angle Neutron Scattering (SANS)

(i) *Unstretched Samples - Isotropic Scattering*

Small angle neutron scattering experiments were also carried out to obtain information on clay structure within the nanocomposite. In contrast to TEM that images a 50 nm-thick sample over micron-sized field, SANS experiments assess clay structure over a ~20 mm$^3$ sample volume (illumination area $\cong$ 20 mm$^2$, sample thickness $\cong$ 1 mm). Pristine rubber shows strong scattering due to non-rubber particles, the most probable candidate of which being calcium sulfate crystallites that are formed in dry films. Consequently, scattering signal from clay particles in the nanocomposite films is masked or depressed. This excess scattering from non-rubber components is significantly reduced in dialyzed films, where although some scattering remains, its intensity is much lower compared to that of the clay particles. The isotropic scattering curves of dialyzed nanocomposites (DNR) at different clay content are shown in Figure 6. These curves have been normalized by the rubber volume fraction (1-$\phi$), and the convergence of the spectra at high-q is due to incoherent scattering from the hydrogenated rubber matrix.

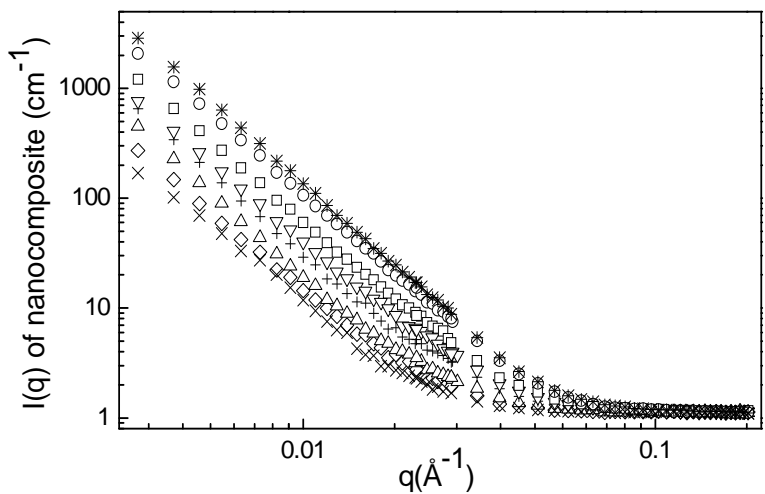

Figure 6. Scattering from DNR nanocomposites at different clay concentrations: 0% (cross), 1% (diamond), 2% (triangle), 4% (plus), 6% (inverted triangle), 10% (square), 20% (circle), 30% (star). Convergence of incoherent scattering at high-q is obtained by normalization of scattered intensity by the rubber volume fraction (1-$\phi$).

Scattering signal from clay particles is obtained by subtracting the spectrum of pure rubber matrix from that of nanocomposite, taking into account the rubber volume fraction (1-$\phi$). A plot of the clay scattering curves is presented in Figure 7. The scattering intensity increases with clay concentration as expected and, in all cases, no small-angle rise is detected in the low q-range indicating absence of large clay aggregates. In this logarithmic representation, theoretically, a slope of -2 is predicted for non-interacting flat discs, as would be the case for completely exfoliated clay platelets. Experimentally however, slopes higher than -2 are commonly obtained, even for aqueous dispersions of clay under dilute and non-interacting conditions, as reported by Hermes et al [34]. The authors attributed the higher slope to presence of tactoids, with the slope increasing with tactoid size. At 1% tactoid with 7 platelets/tactoid for example, their theoretical predictions yield a slope of -2.2. At higher tactoid concentration (10%), 2 distinct slopes are predicted: a $q^{-2}$ platelet component at low-q (~0.001 to 0.03 Å$^{-1}$) and a $q^{-4}$ (Porod) component high-q (>0.03 Å$^{-1}$) arising from surface scattering of the tactoids. The cross-over of these two regions, q*, thus shifts to lower-q value as the tactoid size increases. Slopes of the order of -2 to -3 are typically observed in clay suspensions [34, 35, 36, 37].

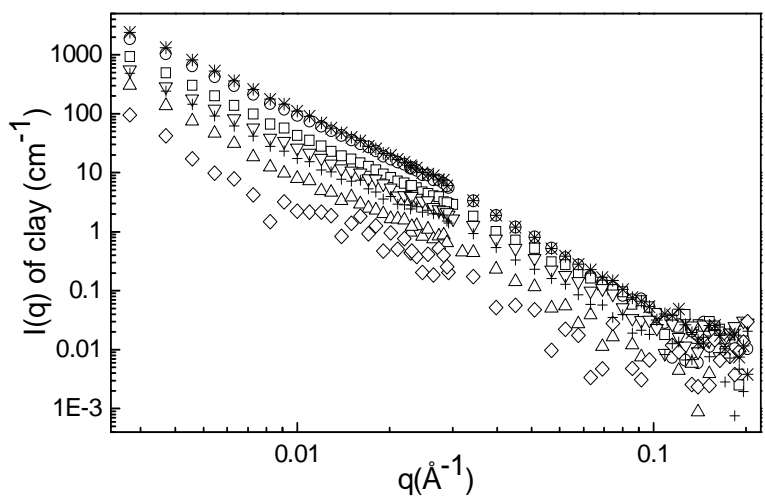

Figure 7. Scattering from clay in DNR nanocomposites at different clay concentrations: 1% (diamond), 2% (triangle), 4% (plus), 6% (inverted triangle), 10% (square), 20% (circle), 30% (star).

The clay scattering spectra in Figure 7 show that for the DNR-clay nanocomposites at clay concentrations higher than 2%, two distinct slopes are discernible, one at low-q (0.003 to 0.02 Å$^{-1}$) and another at high-q (0.02 to 0.2 Å$^{-1}$) region. Linear fits of the different slopes are shown for 6% and 20 % clay in Figure 8. The inset in this Figure shows a clearer way to visualize the slope change by plotting scattered intensity multiplied by q raised to the power of the slope (Iq$^{slope}$) versus q; using this representation, the cross-over q* value can be identified more precisely.

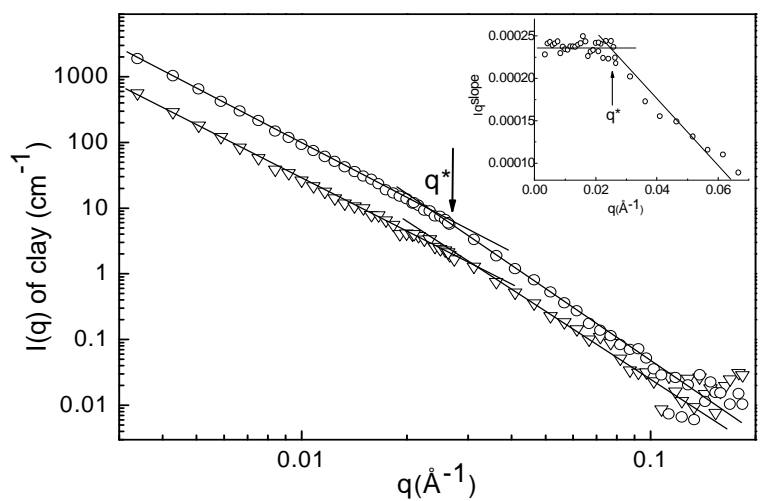

Figure 8. Scattering from clay in DNR nanocomposites containing 6% (inverted triangle) and 20% (circle) clay. Change in slope occurs at q*. Inset: plot of Iq$^{slope}$ versus q shows more clearly the q* value.

The values of the slopes are given in Table 3. It can be seen that for clay concentrations up to 10%, the low-q slopes remain constant at about -2.7, while the high-q slopes increase from -2.6 to around -3.1. At 20 and 30% clay, the corresponding low- and high-q slopes increase to -2.9 and -3.5, respectively. Thus, the Porod-component becomes more marked at these elevated clay concentrations.

Table 3. Slopes of SANS curves of DNR nanocomposites at low- and high-q regions, and the cross-over value q*.

| Sample | Low-q slope | High-q slope | q* |
|---|---|---|---|
| DNR 1 | -2.6 | -2.6 | - |
| DNR 2 | -2.7 | -2.7 | - |
| DNR 4 | -2.7 | -3.2 | 0.027 |
| DNR 6 | -2.7 | -3.0 | 0.026 |
| DNR 10 | -2.7 | -3.1 | 0.023 |
| DNR 20 | -2.8 | -3.5 | 0.023 |
| DNR 30 | -2.9 | -3.4 | 0.024 |

These results indicate that at clay concentrations above 2%, at least 10% of the clay platelets exist in tactoid form. Following to the evaluations of Yoonessi et al [38], the low-q slopes yield about 7 platelets/tactoid for DNR1 to DNR10, and 11-14 platelets/tactoid for DNR20 and DNR30. A very slight corresponding shift of q* to lower value is also perceptible. The structures of clay deduced by SANS are thus consistent with TEM results. The slope from the high-q region also provides information on surface irregularities. Theoretically, in this region, scattering decreases asymptotically as $q^{-4}$ for smooth surfaces (Porod). For non well-defined interfaces, deviation from $q^{-4}$ behavior can be written generally as [39]: $I(q) \sim q^{-(6-D_s)}$ where $D_s$ is the "surface fractal dimension", and $D_s = 2$ for smooth surfaces (Porod) and $2 < D_s < 3$ for rough interfaces. For the nanocomposites studied here, we obtain $D_s = 2.6 - 3.0$, indicative of rough tactoid surface.

(ii) *Stretched Samples - Anisotropic Scattering*

The structural features of clay in the nanocomposite films after stretching were also analyzed by SANS. The stretched sample was clamped to a holder and placed in the neutron beam. Figure 9 shows two-dimensional scattering patterns transformed from isotropic scattering to anisotropic scattering for samples stretched to λ = 4 at different clay concentrations. Corresponding images for pure rubber film show very low scattering resulting from residual non-rubber components. The anisotropic patterns of stretched films show much higher scattering in the horizontal direction (perpendicular to stretching direction) compared to the vertical direction (parallel to direction of stretching). This picture is interpreted as strain-induced reorganization of the scattering objects where distance between objects increases parallel to the stretch direction with a concomitant decrease in distance perpendicular to the stretching direction. This anisotropy increases with clay concentration and with degree of stretching.

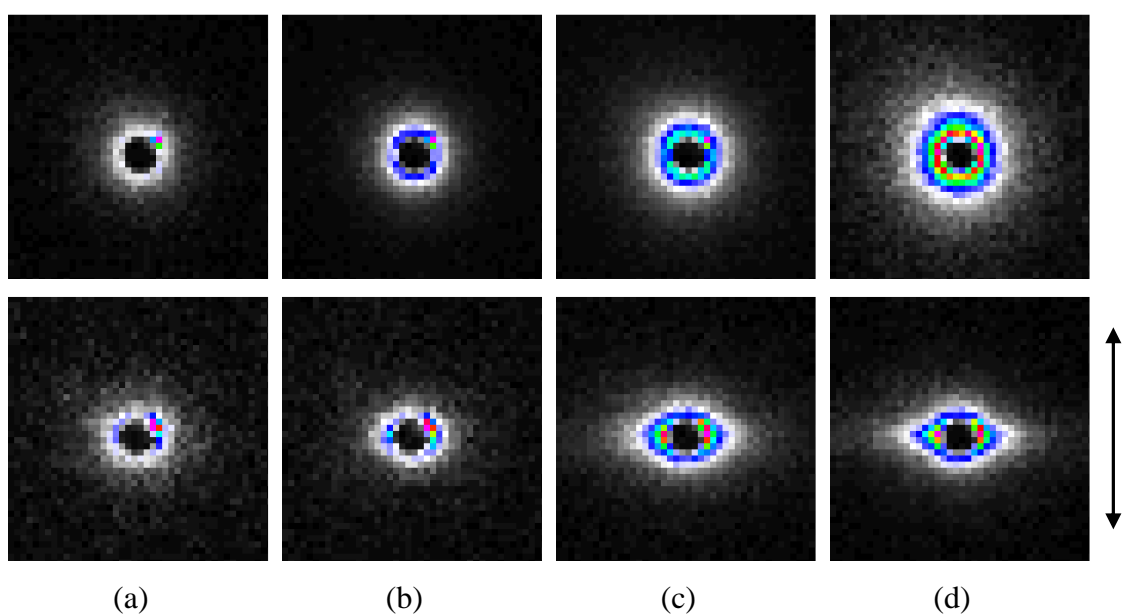

   (a)     (b)     (c)     (d)

Figure 9. Two-dimensional SANS patterns of DNR nanocomposites: non-stretched (top images) and stretched to λ = 4 (bottom images). Clay concentrations: (a) 0%, (b) 2%, (c) 10% and (d) 20%. The vertical arrow shows the stretching direction.

For anisotropic scattering patterns, iso-intensity averaging of horizontal and vertical scattering is performed by considering area sectors ± 10° about the x-axis (horizontal) and ± 20° about the y-axis (vertical). Sector-averaged scattering intensities due to clay particles in stretched samples are obtained by subtracting the signal of pure rubber matrix stretched to the same λ. The resulting scattering spectra for samples stretched to λ = 4 normalized by the rubber volume fraction (1-φ) are presented in Figure 10. As observed in the 2-dimensional image, scattering intensity in the horizontal direction is higher than that in the vertical direction. Note that a difference in horizontal and vertical scattering is observed only at low-q; at high-q, the two scattering curves converge. The slopes for both horizontal and vertical scattering are similar and do not differ from those of non-stretched samples, with $q^{-2.7}$. Thus, uniaxial stretching of the sample results in alignment of clay with respect to the stretching direction without detectable change in clay morphology.

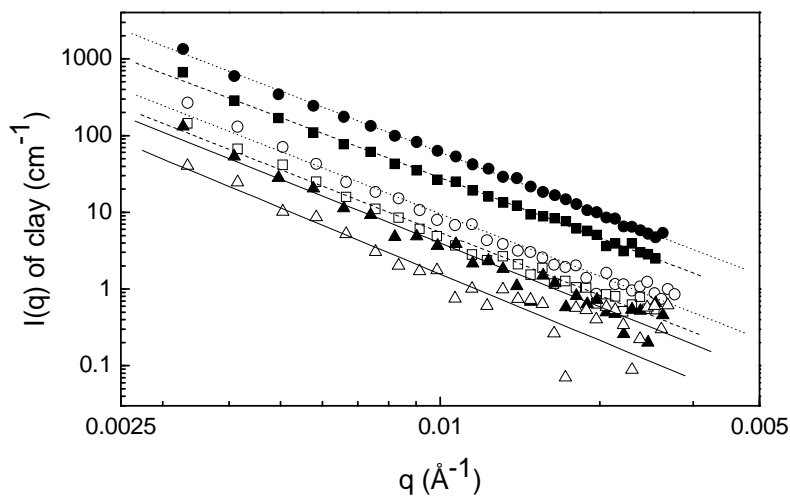

Figure 10. Scattering from clay in DNR nanocomposites stretched to $\lambda = 4$. Horizontal (closed symbols) and vertical (open symbols) scattering are shown for 2% (triangle), 10% (square) and 20% (circle) clay. The lines are linear fits to the experimental data.

## DISCUSSION

In this work, we study the mechanical properties of non-vulcanized natural rubber reinforced by smectic montmorillonite clay for non-dialyzed (NR) and dialyzed rubber (DNR). These properties are examined in relation to the morphological aspects of the clay as determined by TEM and SANS. The rubber-clay nanocomposite films are prepared by mixing aqueous dispersions of pre-exfoliated clays with the natural rubber latex. This is an alternative method that takes advantage of water as an excellent swelling and exfoliating agent for the clay, dispensing the need for chemical modification or organocation exchange.

<u>Pure rubber films</u>

Before discussing the nanocomposites films, some interesting intrinsic properties of natural rubber are to be noted. In the absence of reinforcing agent, non-vulcanized pristine rubber film can be stretched without rupture up to $\lambda_{max} \approx 10$. The stress-strain profile shows an abrupt increase in stress at high deformations, yielding an exceptionally high tensile strength: stress at rupture $\sigma_{max} \approx 5$ MPa. The upturn in the stress-strain profile (for non-reinforced film) is characteristic of strain-induced crystallization of the polymer chains, forming crystalline domains that act as cross-link junctions in the rubber matrix. This crystallization phenomenon has been studied in detail using synchrotron X-rays [40, 41, 42, 43]. In natural rubber, the ease of strain-induced crystallization is generally attributed to the isomeric purity of the polyisoprene from *Hevea brasiliensis* (almost 100% *cis*-1,4 polyisoprene). It is further believed to be promoted by naturally-occurring fatty acids [44, 45]; hence, the superior mechanical properties of natural rubber over its synthetic counterpart. In addition, natural rubber contains a host of inorganic ions. Several of the metallic ions have been identified in our past studies using synchrotron X-ray fluorescence and electron energy-loss spectroscopy imaging in low-energy TEM. In particular, calcium ions are found to be closely associated with gel formation of the polyisoprene matrix [21], and also to induce formation of calcium sulfate crystallites that contribute to hardening in dry aged films [22]. All these factors contribute to auto-reinforcement both in terms of tenacity and tensile strength of the natural rubber.

When rubber latex is dialyzed, the dry film that is formed shows a softer texture even upon aging. In this case, while the maximum tensile strain is unaffected, the corresponding tensile strength is decreased to $\sigma_{max} \approx 1$ MPa. Dialyzed rubber film therefore loses the exceptional high tensile properties conferred by non-rubber components.

Rubber-clay nanocomposites

For these nanocomposites, the mechanism of reinforcement arises from the insertion of high-modulus lamellar particles that are strongly adherent to the rubber latex. Here, following a quantitative model, it has been shown that electrostatic adhesion at the organic/inorganic interfaces may be as strong as covalent binding [19].

In the present study, it is found that the performance of clay as reinforcement agent is more effective in dialyzed rubber: at 10% clay, a 100-fold increase in elastic modulus is obtained for DNR compared to a 5-fold increase for NR. The significant enhancement factor in DNR is related to the excellent degree of exfoliation of clay, with effective aspect ratio $f \approx$ 200-300, while $f \approx$ 50-60 for NR. The value of $f$ for DNR falls within the range of values of single clay platelets ($f$ = 100-500) obtained by TEM for these samples, and are in agreement with other reported results [46, 47]. Note that compared to other types of fillers, the elastic modulus of DNR at 4% clay is of the same order as that containing 33% carbon black [48]. The lower performance of clay for NR can be explained by the high ionic strength and presence of multivalent cations that promote attractions of the clay lamellae. In dialyzed rubber, most of these inorganic ions are removed; the lower ionic strength increases electrostatic repulsion between the clay lamellae, preserving the excellent dispersion of the pre-exfoliated clay platelets.

The effective performance of clay as reinforcing agent is thus better assessed from dialyzed samples. The Guth-Smallwood (Equation 3) and the modified Tsai-Halpin (Equation 5) give good fits to their mechanical properties versus clay concentration, yielding comparable aspect ratios, $f = 220$ (Guth-Smallwood) and $f = 300$ (modified Tsai-Halpin). Note however, that the Guth-Smallwood equation is a purely hydrodynamic model that, due to its parabolic form, is found to fit most experimental data. No meaningful information pertinent to nanocomposite systems can be extracted regarding particle size and specific surface area and particle-matrix adhesion effects, and, in the case of spherical particles, structural information (disperse spheres versus globular aggregated particles with $f = 1$). The modified Halpin-Tsai equation on the other hand, is formulated to capture the mechanical behavior of the nanocomposite with explicit expression in terms of morphology and structural aspects of the filler. Here, the accelerated reinforcement with filler concentration (rapid-rise region) is related to the filler aspect ratio $f$ as well as its maximum volumetric packing parameter, $\phi_m$ (true volume/apparent volume). The "apparent volume" is attributed to frozen or immobilized rubber matrix in the vicinity of the filler, behaving like a solid domain that contributes to reinforcement. For the DNR system, the modified Halpin-Tsai model gives $f = 300$ and $\phi_m = 0.038$. The high $f$ value gives a correspondingly low $\phi_m$ that shifts the rapid-rise transition region to lower filler volume fraction. Since onset of this transition is generally attributed to organization and network formation of the filler, the modified Halpin-Tsai model thus associates to this network formation trapped or immobilized polymer matrix that further contributes to mechanical properties. For spherical particles, network formation takes place around a critical concentration, $\phi_c \approx 0.10$. For anisotropic particles, it depends on the aspect ratio and for $f >> 1$, calculations based on purely entropically-driven phase transition give $\phi_c \approx 3.55/f$ [12]. Using this relation and $f = 300$, we obtain $\phi_c \approx 0.01$. Experimentally, $\phi_c$ can be seen to occur between 0.01 and 0.02, which is almost an order of magnitude lower than expected for spherical particles. The low maximum packing volume, $\phi_m = 0.038$ obtained in this study, coherent with the high degree of exfoliation, is also much lower than those reported for other clay nanocomposites where values between 0.15 and 0.2 have been reported for styrene-butadiene and acrylonitrile-butadiene, both prepared by co-coagulation of clay and latex [8].

Morphological information of the clay particles obtained by both TEM and SANS indicates existence of tactoids together with completely exfoliated clay platelets. Indeed, co-existence of tactoids with exfoliated platelets is a common occurrence in smectic clay systems and has been reported even for dilute aqueous dispersions [34]. For the DNR nanocomposites containing up to 10% clay, the tactoid size averages 7 platelets/tactoid independent of clay concentration (although the tactoid fraction increases). Interestingly, presence of these tactoids does not appear to compromise the excellent reinforcing property conferred by the exfoliated fraction of the clay. At higher clay concentrations (20 and 30%), larger tactoids of about 11-14 platelets/tactoid are also present; these nanocomposite films are too stiff to stretch and their mechanical properties are not evaluated here.

SANS data also indicate rough fractal-like surfaces of the tactoids. This rough surface morphology may explain the propensity of the clay particles to entrap and immobilize polymer matrix that contributes to reinforcement. Differential scanning calorimetry measurements show that although the $T_g$ of the rubber matrix is not affected by the presence of clay particles, the change in specific heat capacity at this transition (expressed in per gram polymer matrix) decreases with clay content: $T_g$ = -62.9°C, $\Delta C_p$ = 0.40 J/g°C (DNR0), $T_g$ = -62.8°C, $\Delta C_p$ = 0.36 J/g°C (DNR10), and $T_g$ = -63.0°C, $\Delta C_p$ = 0.32 J/g°C (DNR30). At the glass transition temperature, therefore, the presence of clay lowers the fraction of the polymer matrix that is fluidified; for DNR30 for example, the degree of fluidization is about 80%. These data support the concept of clay-immobilized polymer matrix contributing to reinforcement of the mechanical properties. (These DSC measurements were carried out by pre-heating the samples to 80°C to erase the thermal history, then cooled to -100°C, and followed by heating the samples at 5°/min).

The elastic modulus discussed above characterizes mechanical strength and stiffness of the nanocomposite at very low deformations. At high deformations, the stress-strain profile and the maximum tensile strain and strength at rupture are important considerations. Enhancement of elastic modulus and tensile strength by fillers means increased stiffness of the nanocomposite which is accompanied by decreased tensile strain or extensibility of the material. When subjected to high deformations, the natural rubber-clay nanocomposites show anisotropic neutron scattering patterns characteristic of alignment of the clay particles. This strain-induced alignment, parallel to the stretching direction, increases with clay concentration and with degree of deformation; no other difference is measured in the scattering intensity profiles. Uniaxial stretching of the sample thus appears to align the clay particles without detectable change in clay morphology (platelet and tactoid distributions). Thus, apart from strain-induced crystallization of the polymer matrix, re-organization and alignment of the filler network may also be associated with increased mechanical properties at large deformations.

## CONCLUSIONS

The mechanical properties of natural rubber-clay nanocomposites have been studied by uniaxial deformations and their results related to the morphological and structural aspects of the clay characterized by TEM and SANS. Non-vulcanized pristine natural rubber shows outstanding mechanical properties. It can be stretched up to ten times its initial length with exceptionally high tensile strength at rupture. This unique property has in the past been attributed to rapid strain-induced crystallization of the isomerically-pure polyisoprene, a process that is further aided by intrinsically-occurring non-rubber components. The present study shows clearly that removal of these non-rubber molecules decreases this auto-reinforcement effect.

For the natural rubber-clay nanocomposites, the alternative preparation method of "latex mixing" produces macroscopically homogeneous films with excellent degree of exfoliation and dispersion of clay in dialyzed rubber. For non-dialyzed rubber however, the high ionic strength and presence of multivalent cations promote attractions of clay lamellae resulting in lower degree of exfoliation. For the dialyzed sample, the effective aspect ratio of the exfoliated clay falls within the expected range for single clay lamella. Correspondingly, onset of accelerated stiffening of the nanocomposite, attributed to filler network formation, occurs at a very low critical clay concentration that is almost an order of magnitude lower than would be expected for spherical particles. Morphologically, results from TEM and SANS reveal co-existence of completely exfoliated clay platelets with tactoids. Interestingly, presence of these tactoids does not appear to compromise the excellent reinforcement property conferred by the exfoliated platelets. The mechanical properties, together with data from SANS and calorimetry also advocate occurrence of immobilized polymer matrix contributing to mechanical reinforcement.

At high deformations, strain-induced alignment of clay particles is clearly demonstrated by the anisotropic neutron scattering patterns, with the degree of anisotropy increasing with strain and clay concentration. Under these conditions, apart from ordering, no change in the clay morphology is detected. Thus, enhanced mechanical properties at high uniaxial deformations may also be related to ordering of the clay network in the nanocomposite.

ACKNOWLEDGEMENTS

We thank Jacques Jestin for helpful discussions and assistance with the traction machine and with neutron scattering experiments.